\documentclass[reprint,floatfix,aps,twocolumn, noeprint,superscriptaddress,notitlepage]{revtex4-2}
\usepackage{graphicx} 
\usepackage{amsmath}
\usepackage{amsfonts}
\usepackage{amssymb} 
\usepackage{xcolor}
\usepackage{multirow}
\usepackage{lineno}
\usepackage{url}
\usepackage{adjustbox}
\usepackage{setspace}
\usepackage[T1]{fontenc}
\usepackage{booktabs,makecell,eqparbox}
\usepackage{comment}
\usepackage{ifthen}

\begin{document}

\title{Divergent Clinical Equivalence Findings from DVH and NTCP Metrics for Alternative OAR Delineations with Increasing Setup Variability in Head-and-Neck Radiotherapy}

\author{M.N.H.~Rashad} \email[Email: ]{hashir@virginia.edu} \affiliation{University of Virginia, Charlottesville, VA, 22903}
\author{Abishek Karki} \affiliation{University of Virginia, Charlottesville, VA, 22903}
\author{Jason Czak} \affiliation{University of Virginia, Charlottesville, VA, 22903}
\author{Victor Gabriel Alves} \affiliation{University of Virginia, Charlottesville, VA, 22903}
\author{Hamidreza Nourzadeh} \affiliation{Thomas Jefferson University, Philadelphia, PA, 19107}
\author{Wookjin Choi} \affiliation{Thomas Jefferson University, Philadelphia, PA, 19107}
\author{Jeffrey V Siebers} \affiliation{University of Virginia, Charlottesville, VA, 22903} 

\date{\today}

\begin{abstract}
\textbf{Purpose:} 
This study quantifies the variation in dose-volume histogram (DVH) and normal tissue complication probability (NTCP) metrics for head-and-neck (HN) cancer patients when alternative organ-at-risk (OAR) delineations are used for treatment planning and for treatment plan evaluation. We particularly focus on the effects of daily patient positioning/setup variations (SV) in relation to treatment technique and delineation variability.

\textbf{Materials and Methods:}
We generated two-arc VMAT, 5-beam IMRT, and 9-beam IMRT treatment plans for a cohort of 209 HN patients. These plans incorporated five different OAR delineation sets, including manual and four automated algorithms. Each treatment plan was assessed under various simulated per-fraction patient setup uncertainties, evaluating the potential clinical impacts through DVH and NTCP metrics.

\textbf{Results:}
The study demonstrates that increasing setup variability generally reduces differences in DVH metrics between alternative delineations. However, in contrast, differences in NTCP metrics tend to increase with higher setup variability. This pattern is observed consistently across different treatment plans and delineator combinations, illustrating the intricate relationship between SV and delineation accuracy. Additionally, the need for delineation accuracy in treatment planning is shown to be case-specific and dependent on factors beyond geometric variations.

\textbf{Conclusions:}
The findings highlight the necessity for comprehensive quality assurance programs in radiotherapy, incorporating both dosimetric impact analysis and geometric variation assessment to ensure optimal delineation quality. The study emphasizes the complex dynamics of treatment planning in radiotherapy, advocating for personalized, case-specific strategies in clinical practice to enhance patient care quality and efficacy in the face of varying SV and delineation accuracies.
\end{abstract}

\maketitle

\section{Introduction}
Organ-at-risk (OAR) delineations used for radiotherapy treatment planning are assumed to represent the true underlying structure. 
However, inter-observer, intra-observer, and inter-algorithm OAR delineation variations~\cite{VINOD2016, Bhardwaj2008, Xu2015, FIORINO1998, Nourzadeh2017} indicate that clinical delineations are not absolutely accurate. 
Nonetheless, decades of successful radiotherapy have shown that absolute accuracy is not required. 

OAR delineation accuracy requirements depend on factors such as proximity to the target, treatment technique, and the OAR dose-response characteristics.
Additionally, inter-treatment patient setup variability and organ motion/deformation affect the OAR dose, hence influencing the required delineation fidelity.  

Comparisons of alternative manual delineations (MDs) in standardization studies~\cite{Caravatta2014} and between MDs and auto-delineations (ADs) based on geometric indices~\cite{Yang2017, Raudaschl2017, Doolan2023} do not assess their adequacy for treatment planning. Some studies quantify the dosimetric effect of alternative delineations post-planning~\cite{ Andrea2014, Martin2015}, but post-planning evaluation does not evaluate their suitability for treatment planning.  Delineation variations can significantly impact patient treatment~\cite{CARDENAS2019185}.
 
Recent investigations recognize the need to utilize alternative test delineations in the treatment planning process, and
explore correlations between geometric indices and dosimetric variations, revealing complex and case-specific relationships.~\cite{Poel2021, Cao2020, Lim2019} 
Some studies~\cite{Smolders2023, Rooij2019} demonstrate adequacy of ADs for treatment planning; 
others~\cite{Fung2020, Guo2021} find substantial dose differences despite minor geometric variations.
These studies generally utilized few (10-20) patients, a single treatment planning technique, and/or few (e.g. 2) alternative delineation sets.  

This study examines if/how daily patient setup variations affect the clinical impact of alternative delineations in radiation therapy planning.
Our work builds on previous approaches by using a large (209) patient cohort, five alternative delineations, three treatment planning techniques.  We examine the interplay between delineation variability and daily patient setup variations,  
evaluating potential clinical effects using $D_{\text{max}}$, $D_{\text{mean}}$, and normal tissue complication probability (NTCP).   
This comprehensive approach aims to understand how these factors collectively influence the clinical impact of alternative delineations in radiation therapy planning.

\section{Materials and method}
\label{sec:M&M}
For each patient in a 209 head and neck (HN) patient dataset, 2-arc VMAT, 5-beam and 9-beam IMRT treatment plans were created using five alternative OAR sets (one MD, four AD)  using an unsupervised auto-planning algorithm. 
The same MD targets were used for all treatment plan optimizations. 
Each treatment plan was evaluated with each alternative OAR set under six different patient setup uncertainty scenarios.
The potential clinical impact of using alternative structure sets was assessed using DVH and NTCP plan quality indices ($PQI$).
The $\Delta PQI$ between planning and alternative OAR set evaluations were 
compared to clinical tolerances, below which the delineations are considered equivalent.
Differences and similarities in the effects of increasing setup variation on equivalence for $D_{\text{mean}}$, $D_{\text{max}}$, and NTCP $PQI$s for the different OARs and beam arrangements were evaluated.

\subsection{Data curation}
Two-hundred nine HN datasets from two The Cancer Image Archive (TCIA) collections were used in this study. 
Seventy three were from Head-Neck Cetuximab collection~\cite{Bosch2015, Kian2014}, 
and 136 were from Head-Neck-PET-CT~\cite{Vallieres2017TCIA, Vallieres2017} collection.

Most CT images and manual OAR delineations (187) were sourced from the UaNet Github repository~\cite{UaNet_Git}, which curated the delineations~\cite{Tang2019}. 
UaNet Dataset 2 includes 140 CT scans from the TCIA Head-Neck Cetuximab~\cite{Bosch2015} 
and Head-Neck-PET-CT~\cite{Vallieres2017TCIA} collections, with 
up to 28 OARs per patient re-delineated by a single experienced radiation oncologist and reviewed by a second expert~\cite{Brouwer2015, Tang2019}. 
One dataset 2 patient was excluded due to miss-alignment of the PTVs with the CT image set.
UaNet Dataset 3 from the  Public Domain Database for Computational Anatomy (PDDCA - Version 1.4.1), includes 48~CTs with up-to 9 manually delineated OARs from the Head-Neck Cetuximab collection~\cite{Bosch2015} which were re-segmented for use in the 2015 Head and Neck Auto Segmentation MICCAI Challenge~\cite{Raudaschl2017}. The remaining 22 patients were processed in-house from the Head-Neck-PET-CT~\cite{Vallieres2017TCIA} collection. For all patients, PTVs were selected from the TCIA collections.Patients in the dataset were limited to those from which we could discern an unambiguous association between the CTs and the corresponding aligned contour sets.

Auto-delineations  for each CT image set were created using AutoContour (Radformation Inc~\cite{Radformation}), INTContour (Carina Inc~\cite{Carina}), Syngo.via (Siemens Healthineers) and SPICE (Pinnacle, Philips Professional Healthcare), referred to as AD1, AD2, AD3, and AD4.
All auto-delineations were used without modification to ensure delineation variability, with grossly erroneous delineations eliminated through geometric comparisons. 

\begin{figure}[htbp]
\begin{center}
\includegraphics[width=0.95\linewidth]{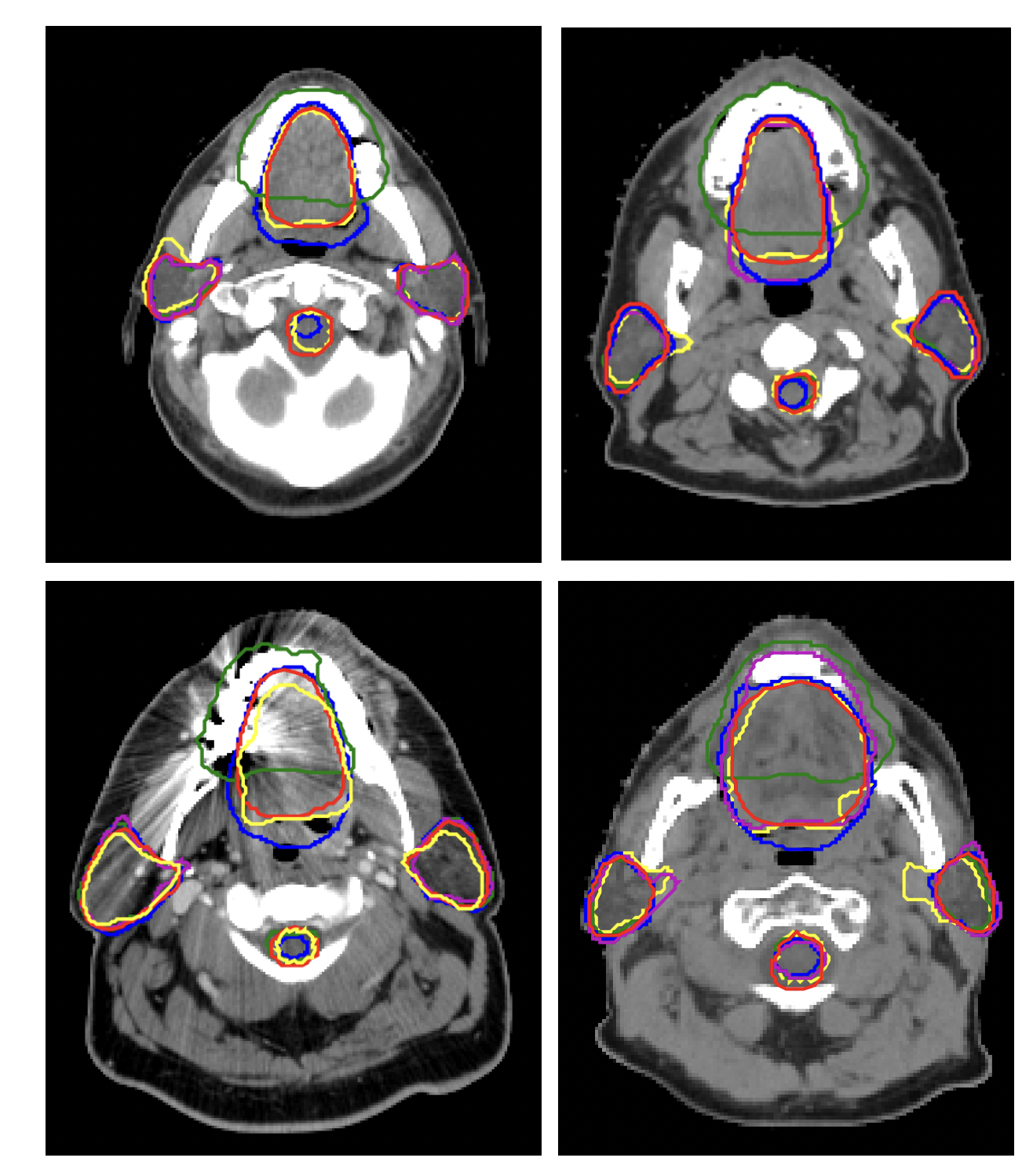}
\end{center}
\caption{Examples of the variability observed between the alternative delineations (manual and auto) for four different patients.  Only SpinalCord, Parotid\_L, Parotid\_R and Cavity\_Oral are shown for clarity.
The contours colors are, 
MD=Manual: Magenta, 
AD1=Radformation: Red, 
AD2=Carina: Green,
AD3=Siemens: Blue and 
AD4=SPICE: Yellow.
The observed DV is patient and OAR specific.}
\label{fig:DV}
\end{figure}

Figure~\ref{fig:DV} compares the alternative delineations of four OARs for four patients illustrating variations in the alternative delineations. 

\subsection{Geometric comparisons}

Volumetric Dice Similarity Coefficient (vDSC) and robust Hausdorff Distance (HD95) geometric indices (GIs) were computed for all OAR and delineator combinations using methods from Alphabet Inc. Google DeepMind~\cite{gDeepMind_SD}. 
GIs were used for first-order delineation quality assurance and to evaluate correlations between geometric and dosimetric differences. 
For dosimetric and NTCP analysis, delineation pairs with vDSC<0.5 were excluded. 
For the SpinalCord, geometric comparisons were limited to CT slices common between both delineators, which we define as "common-slice-Dice".

\subsection{Treatment plan creation}

Two-arc VMAT, 5-beam IMRT and 9-beam IMRT plans were created for each patient using the MD targets and each alternative structure set, totaling 15 plans per patient (3135 plans in total) with the auto-planning algorithm in Pinnacle 16.2. 
All patients used the same base PTV prescription dose-levels, 70, 63 and 56 Gy in 35 fractions, regardless of the clinical plan dose level.
For Patients with <3 PTVs,  the prescription limited to the highest dose level PTVs. 
Using higher-than-clinical dose levels was a conservative approach 
as it results in higher OAR doses and higher sensitivity to delineation variations.
Each plan was optimized for the same base objectives (Table~\ref{tab:NTCPParameters}). 
OARs not present in a delineation set were excluded from that optimization. 
Differences in optimization OAR set minimally affected the validity delineation equivalence assessments (section~\ref{sec.AlternativeTruthMethod}), as the missing OAR was then excluded from the pair-wise analysis. 

\begin{table}[htbp]
\centering
\begin{adjustbox}{width=\linewidth}
\begin{tabular}{c|ccccc|cccc}
\toprule
\multicolumn{1}{c}{} &
\multicolumn{5}{c}{\textbf{Optimization Objectives}} &
\multicolumn{4}{c}{\textbf{NTCP Parameters}} \\	
\cmidrule(lr){2-6}\cmidrule{7-10}
     
\makecell{\textbf{OAR Name}} &
\makecell{Dmax \\ (Gy)} &
\makecell{Dmean \\ (Gy)} &
\makecell{DVH\_V \\ (Max \%)} &
\makecell{DVH\_D \\ (Max Gy)} &
\makecell{Priority} &
\makecell{n} &
\makecell{m} &
\makecell{$TD_{50}$}&
\makecell{End Point}\\
\midrule
BrachialPlexs (L\&R) &  66 &  - &  3 &  62  & Low & - & - & -& -\\
Brainstem &  54 &  - &  5 &  52 & High & 0.16 & 0.14 & 65 & Necrosis/infraction\\
Glnd\_Submands (L\&R) &  - &  35 &  - &  - & High  & 0.70 &  0.18 &  46, 56 &  Xerostomia\\
Larynx &  63 &  - &  3 &  39 & High & - & - & -& -\\
Bone\_Mandible &  75 &  - &  - &  - & High &  0.07 &  0.10 & 72 &  Marked limitation of joint function\\
OpticChiasm &  44 &  - &  - &  - & High & - & - & -& -\\
OpticNrv (L\&R) &  55 &  - &  - &  - & High & - & - & -& -\\
Parotid (L\&R) &  - &  26 &  7 &  20 & High & 0.70 &  0.18 &  46 &  Xerostomia\\
SpinalCord &  45 &  - &  - &  - & High &  0.05 &  0.175 &  66.5 & Myelitis/necrosis\\
Trachea &  69 &  - &  5 &  60 & High & - & - & -& -\\
\bottomrule
\end{tabular}
\end{adjustbox}
\caption{ Describes the optimization objectives used in the study by the auto-planning algorithm and NTCP parameters, n, m and $TD_{50}$ used in the study. These parameters (except for Glnd\_Submands) are obtained from~\cite{Burman1991}. For Glnd\_Submands, we used the same parameters as the Parotid as well as a slightly elevated $TD_{50}$.
}
\label{tab:NTCPParameters}
\end{table}

\subsection{Setup variability (SV) simulation - RTRA}
\label{sec.RTRA}
To understand the effects of alternative delineations amidst inherent treatment uncertainties, including patient setup uncertainties, we used the Radiation Treatment Robustness Analyzer (RTRA)~\cite{Nourzadeh2017} to simulate the impact of setup uncertainties on the planned dose distribution, dose-metrics, and normal tissue complication probability (NTCP). 

RTRA simulates OAR setup uncertainties using rigid body translations in the left-right, anterior-posterior, and superior-inferior directions. These translations are sampled from zero-centered normal distributions with user-set standard deviations for random (per fraction) ($\sigma$) and systematic (per treatment course) ($\Sigma$) uncertainties.  
The translated OARs are combined with the planned dose to evaluate dose volume histograms (DVH) and dose volume coverage map (DVCM)~\cite{Gordon2010} for 1000 treatment course simulations.
The DVHs and DVCMs are then used to compute probabilistic $PQI$s, assessing the probability of achieving a given $PQI$, with evaluations at the 95\% confidence level. 

To assess the effect of setup uncertainty on the $PQI$s, we simulated setup uncertainties with $\sigma = \Sigma \in [0, 2, 4, 6, 8, 10] \text{ mm}$. These values range from the static plan (0 mm) to various clinical setup uncertainties, including IGRT-based setups, laser-based setups, and extending to clinically unrealistic large uncertainties.  

The largest simulated uncertainty, while extending beyond typical clinical scenarios, 
enabled us to quantify trends in the $PQI$ assessments. 
This comprehensive analysis helped determine if and when setup variability outweighs delineation uncertainty, providing an understanding of whether permissible delineation variability depends on the setup uncertainty. 

\subsection{Normal tissue complication probability (NTCP)}

For a given OAR, delineator, and simulated setup uncertainty, NTCPs were computed using the Lyman-Kutcher-Burman (LKB) model~\cite{Burman1991} for each of the 1000 treatment course simulations per SV level. Table~\ref{tab:NTCPParameters}) lists the $n$ (volume-effect parameter), $m$ (dose-response slope) and $TD_{50}$ (uniform irradiation dose resulting in 50\% complication) used from~\cite{Burman1991}.

\begin{equation}
    NTCP = \frac{1}{\sqrt{2\pi}}\int_{-\infty}^{t} e^{\frac{-x^2}{2}} dx
\end{equation}

\begin{equation}
    t = \frac{EUD - TD_{50}}{m TD_{50}},
\end{equation}

with, equivalent uniform dose $EUD$ equal to the generalized mean dose $gMD$~\cite{NIEMIERKO1991} computed from
RTRA computed differential DVHs dose-volume pairs $\{D_i, v_i\}$for each treatment course Niemerko's DVH reduction scheme~\cite{NIEMIERKO1991}.

\begin{equation}
    EUD = gMD= \Bigg(\sum_{i} v_{i}D_{i}^{1/n}\Bigg)^n
\end{equation}

Although the analysis was performed for all organs listed in Table~\ref{tab:NTCPParameters}, results are presented only for the SpinalCord, (a serial organ whose response is proportional to the maximum dose), and the Parotids, (a parallel organ whose response is proportional to the mean dose).  
Parotid glands were separated into those intersecting a PTV and those that do not.

\subsection{Delineation equivalence assessment}
\label{sec.AlternativeTruthMethod}
For brevity, we define the planning delineation (PD) as the delineation set used for the treatment plan creation, and test delineation (TD) as the delineation set used for plan evaluation.
When the TD-based evaluation meets the plan objectives, the PD OARs were adequate for the task of plan creation  when TD represents the true underlying organ.  

We quantify delineation equivalence by the difference in the $PDI$ between the plan evaluated with the same structure set (PD($A$)) used for plan creation ($PQI_{AA}$) (PD$=$TD) and the plan evaluated with an alternate structure set B ($PQI_{AB}$) (PD$\ne$TD). 

TD($B$) is clinically equivalent to PD($A$) if 
\begin{equation}
\label{eq.Delta_PQI}
\Delta PQI_{AB} = |PQI_{AA}-PQI_{AB}| < C_{Tol}
\end{equation} 
where $C_{Tol}$ is the clinical tolerance. 

Reversing the roles of PD and TD structure sets (structure set $B$ used for plan creation and structure set $A$ for plan evaluation) evaluates $\Delta PQI_{BA}$, the clinical equivalence of TD($A$) with PD($B$).

Since reversing these roles results in a different optimized dose distribution, 
generally, $PQI_{AA} \neq PQI_{BB}$, 
$PQI_{AB} \neq PQI_{BA}$, and $\Delta PQI_{AB} \ne \Delta PQI_{BA}$.
Hence, joint equivalence of $A$ and $B$ requires,

\begin{equation}
\label{eq.ClinicalEquivalency}
    \text{Max}(\Delta PQI_{AB}, \Delta PQI_{BA}) < C_{Tol}
\end{equation}

While a two-way assessment is necessary to establish full delineation equivalency, a one-way assessment is sufficient to demonstrate that plans created with PD are adequate when TD represents the underlying organ,  
even though the reversal of the PD and TD may result in clinical non equivalency. 

Delineation equivalence assessments were computed for each (TD, PD) pair for each treatment plan. 
With 5 delineators, we performed 4 assessments per PD and 20 one-way assessments in total per plan. 
Considering 3 treatment planning techniques, we have 60 total one-way assessments per patient for each level of setup uncertainty. 
Assessments without considering the effects of setup variability (equivalent to $\sigma = \Sigma = 0$) and those including setup variability were computed.

\section{Results}

\subsection{Relationship between geometric and dosimetric variations} 

\begin{figure}[htbp]
\begin{center}
\includegraphics[width=\linewidth]{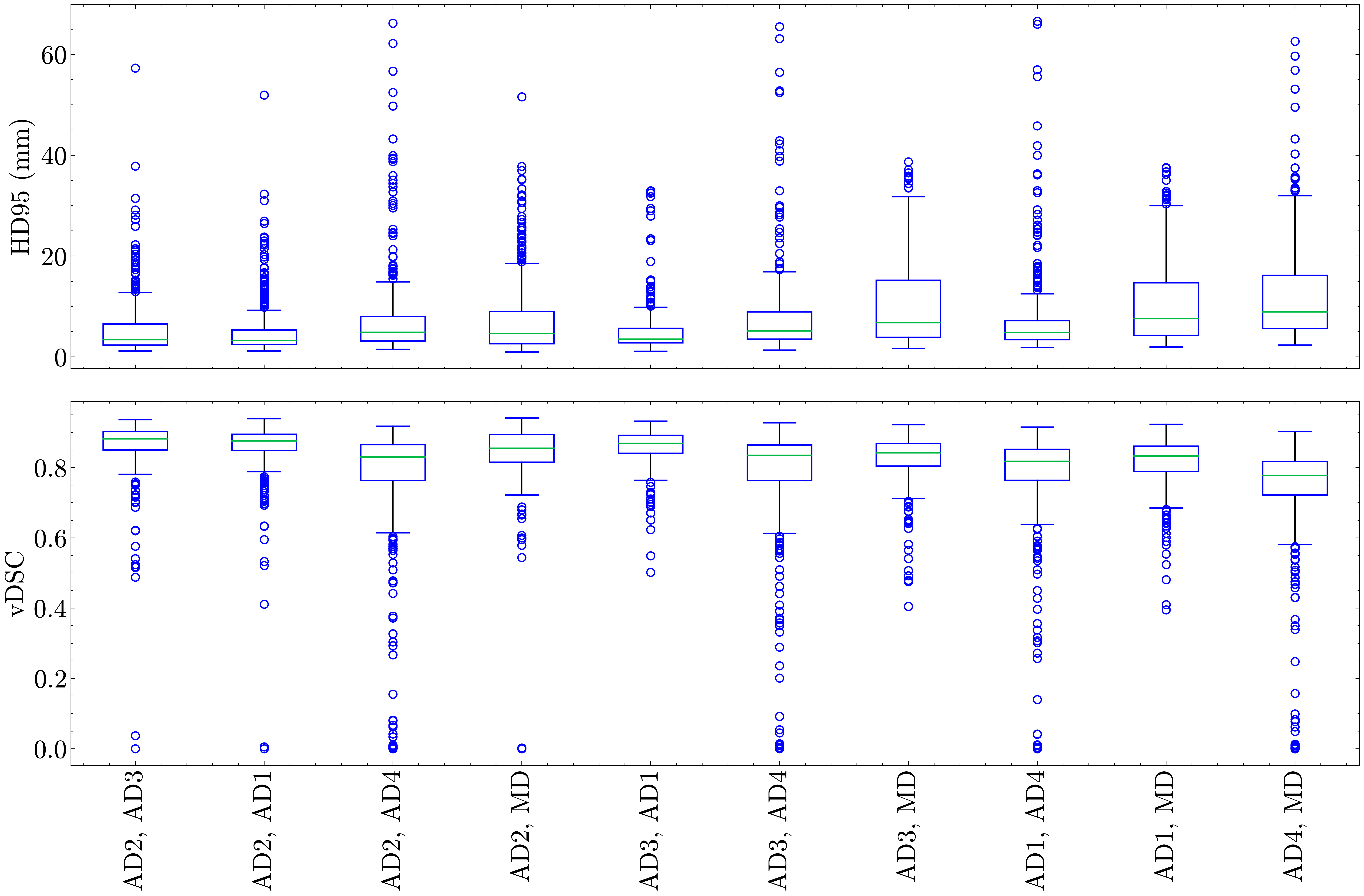}
\end{center}

\caption{Distribution of HD95 in mm (top) and vDSC (bottom) for Parotids (Left + Right) for all of the delineator combinations.  The vast majority of delineations had vDSC greater than 0.8. Note, cases with vDSC<0.5 were excluded from the dosimetic analysis. Geometric differences were greatest for comparisons involving delineator AD4, a model-based auto-delineation method. 
}
\label{fig:ParotidGI}
\end{figure}

Figure~\ref{fig:ParotidGI} shows the vDSC HD95 values for the combined Parotids (Parotid\_L + Parotid\_R) for each PD-TD pair.
All structures and delineator pairs had a median HD95$\leq$0.8 mm and median vDSC$\geq$8. 
With the exception of the AD4 contours, few delineations had vDSC$\leq$0.5 or HD95$\geq$15 mm. 
For delineations with vDSC$\leq$0.5, one delineation in the comparison pair is labeled erroneous. 

\begin{figure}[htbp]
\begin{center}
\includegraphics[width=\linewidth]{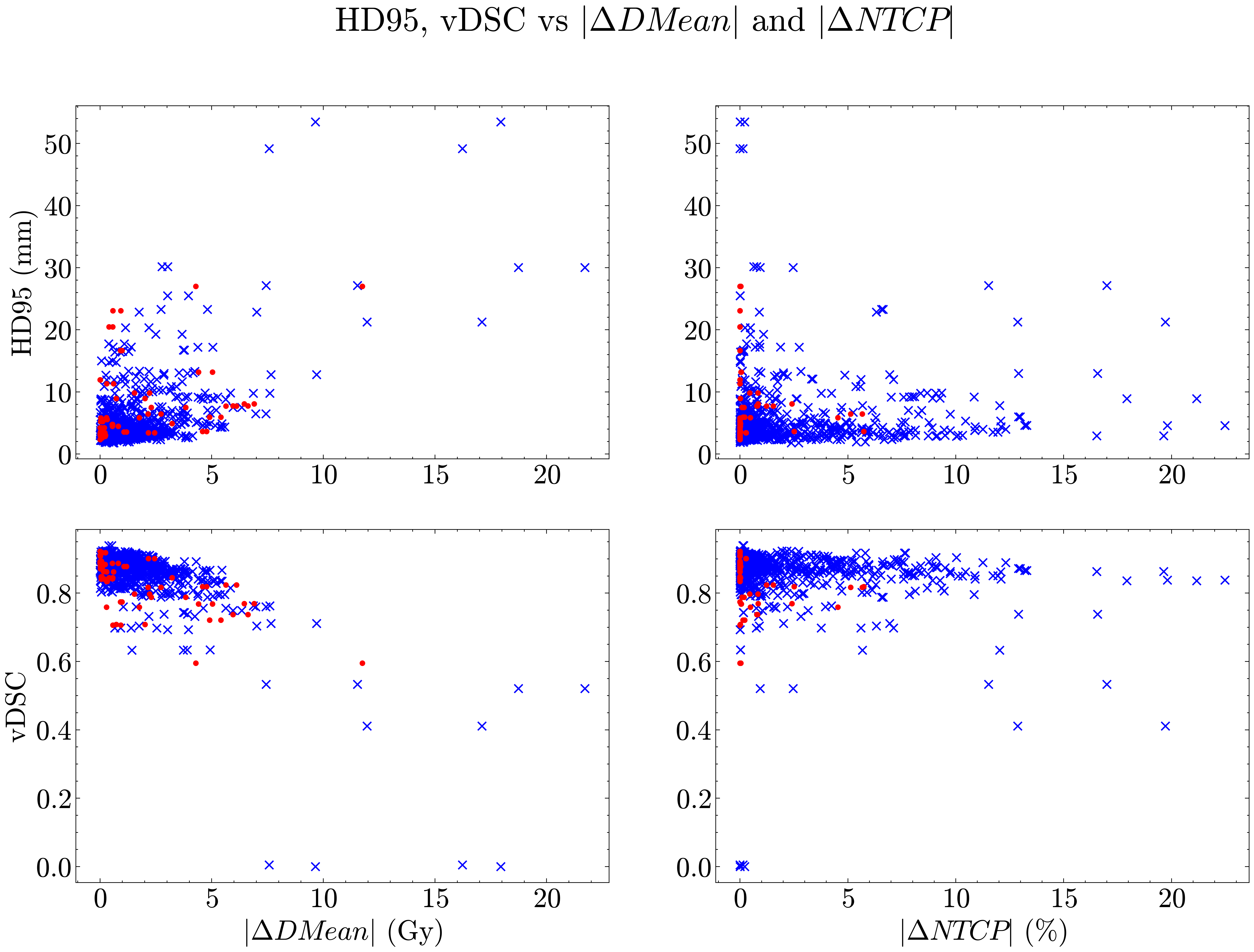}
\end{center}
\caption{Relationship between the geometric indices for alternative Parotid (L+R) delineations and the $\Delta D_{\text{mean}}$(Gy) 
(left column) and $\Delta NTCP$ (\%) (right column) for AD1 and AD2 combinations for all patients for the 2arc VMAT plans. Top row: HD95, bottom row: volumetric DSC. The $\Delta$'s were evaluated for the static plans (no SV simulated). Parotids which overlap with a target volume are shown in blue.  Those with overlap, are in red.}  
\label{fig:GeometryVsDosimetry}
\end{figure}

The correlation of geometric indices (GI) with $\Delta PQI_{AB}$, excluding the effect of SV (for $\Sigma=\sigma=0\text{ mm}$), is shown in 
Figure~\ref{fig:GeometryVsDosimetry} for Parotids' $\Delta D_{\text{mean}, AB}$(Gy) and $\Delta NTCP_{AB}$ (\%) for the combination of AD1 and AD2 delineations with the 2arc VMAT plans. 
While weak correlation are discernible, large variations in $\Delta D_{\text{mean}}$(Gy) and $\Delta NTCP$(\%) for the same GI, along with the existence of small dose and $NTCP$ deviations despite large geometrical differences suggests that the delineation accuracy required for treatment planning is case-specific and depends on factors beyond simple geometrical variations. 
Similar weak correlations (not shown) are observed for other OARs
and for the 5- and 9-beam plans.
This indicates that a comprehensive delineation QA program should consider dosimetric impact analysis in addition to geometrical variation analysis.

\subsection{Effect of setup variability on clinical impact of DV}

To assess the effect of setup variability on delineation variability, we present the results of the union of one-way assessments between PD, TD pairs ($\Delta PQI = \Delta PQI_{AB} \cup \Delta PQI_{BA}$).
Our overall remain consistent whether we consider one-way assessments, the union of one-way assessments, or two-way assessments. 
While we restricted our presented results to PD-TD pairs which have vDSC$\geq$0.5, including vDSC<0.5 yields similar observations as discussed below.
 
\subsubsection{\textbf{Qualitative assessment}}

Equation~\ref{eq.ClinicalEquivalency} outlines the criterion used to evaluate the clinical equivalency of alternate delineations. 
Instead of using a fixed $C_{Tol}$  value,  we analyse the cumulative distribution function (CDF) of $\Delta PQI$ to assess equivalency as a function of $C_{Tol}$, demonstrating the robustness of our findings to $C_{tol}$.  
 
\begin{figure}[htbp]
\begin{center}
\includegraphics[width=\linewidth]{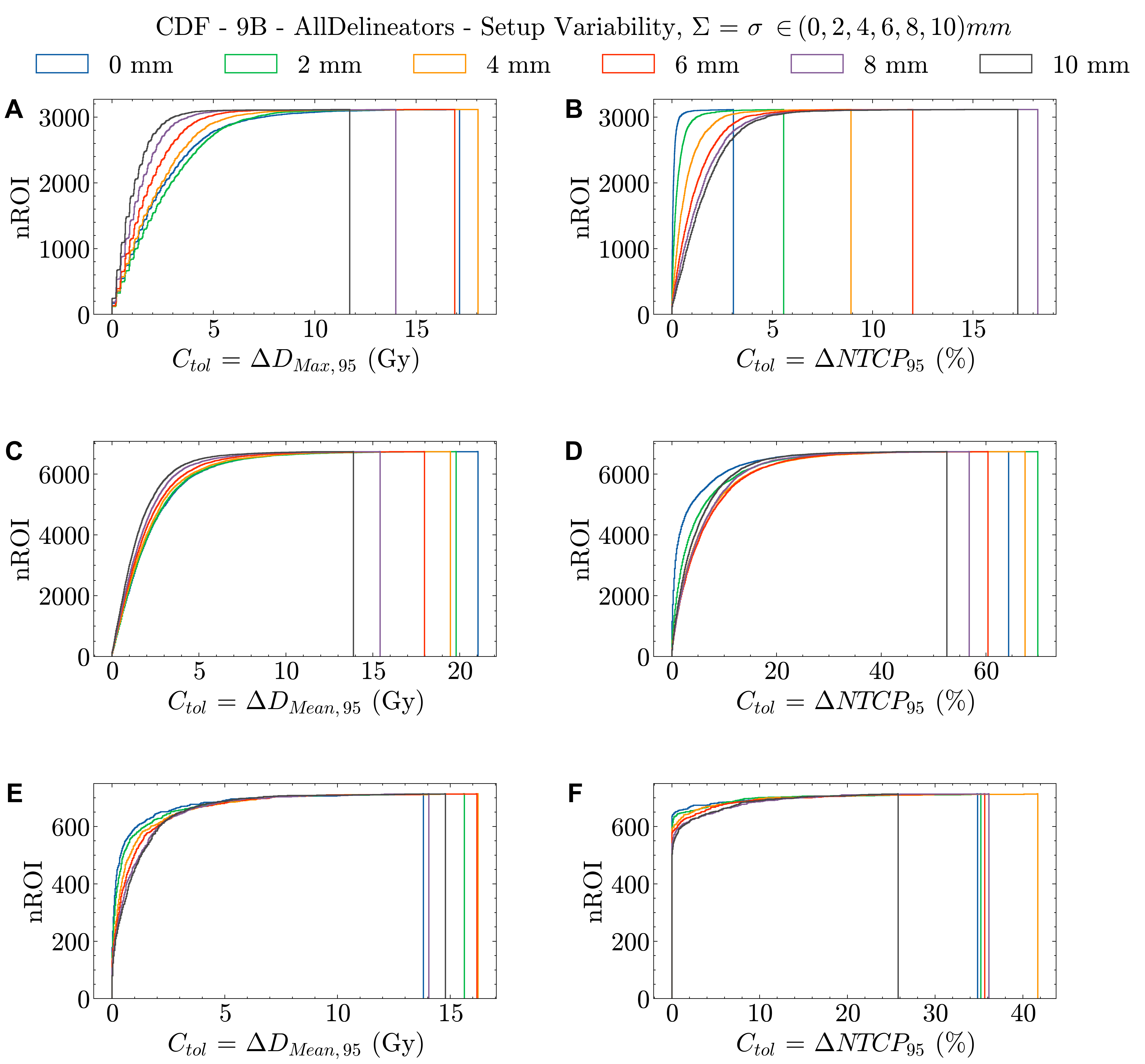}
\end{center}

\caption{Cumulative histograms of the union of $\Delta PQI_{AB}$ for $A, B \in {MD, AD1, AD2, AD3, AD4}$ with $A\neq B$, showing the number of delineations with $\Delta PQI$ < $C_{tol}$ for different simulated setup variabilities (SV).
$\Delta PQI$s are evaluated at the 95\% confidence level. 
Each data series shows the number of equivalent delineations (A) CDF of SpinalCord $\Delta D_{Max, 95}$, indicating an increase in equivalent delineations with increasing SV. 
(B) CDF of SpinalCord $\Delta NTCP_{95}$, showing an 
 increase in  $\Delta NTCP_{95}$ with increasing SV. 
(C) and (D) show similar trends for Parotid Glands overlapping with the target 
(E) and (F), however, show a decrease in equivalent delineations with increasing for non-overlapping Parotids.
The plot is for 9-beam IMRT plans, consistent with 5-beam and 2-arc plans.} 
\label{fig:DVVSSV}
\end{figure}

Figure~\ref{fig:DVVSSV} compares
the alternative delineations, including the effect of varying amounts of SV for SpinalCord (Serial - MaxDose organ) and Parotids (Parallel - MeanDose organ) as evaluated by $\Delta DVH_{95}$ and $\Delta NTCP_{95}$ metrics. Each series shows  the behavior of $\Delta PQI$ as a function of the simulated SV. 
On the CDFs, the Y value at a given $\Delta PQI=C_{tol}$ on the X axis indicates the number of in-tolerance (equivalent) delineations. Conversely, for a given nROI on the Y, the X value gives the associated $C_{tol}$. This enables us to infer the behavior of equivalency at a series of tolerance values.

The observations can be summarized as follows,

\begin{itemize}
\item
Increasing SV generally increases the number of equivalent OARs for all DVH Metric evaluations, indicating a washout effect, except for non-overlapping parotids, where increasing SV decreased the number of equivalent OARs.  
\item
Increasing SV decreases the number of equivalent alternative OARs decreases when evaluated using NTCP for all OARs except for Glnd\_Submands, which washout effects for both DVH and NTCP. 
\item
These trends were consistent across all planning techniques and delineator combinations studied.
\end{itemize}

These differences in the effect of SV on the impact of DV, as measured by DVH Metric vs NTCP, suggests that DVH Metrics may be poor proxies for clinical effect, similar to the findings of 
~\cite{DVHEffect_Lung}. 
This supports the TG 166 recommendation to use biologically related models for treatment planning~\cite{TG166, TG166_shortReport} and highlights the need to consider an 
endpoint metric such as NTCP in any dosimetric impact analysis to determine  required delineation quality.

\subsubsection{\textbf{Quantitative assessment}}
The CDFs in Figure~\ref{fig:DVVSSV} are from $\Delta PQI_{i}$ evaluations  for each $i \in 2 \times n\_ROI$s from the $A=MD$, $B=AD1$ evaluations. 

Defining the equivalency fraction as
\begin{equation}
\label{Feq}
    F_{eq} = \frac{nROI(\Delta PQI \leq C_{tol})}{nROI_{total}}
\end{equation}
allows evaluation of the $C_{tol}$ to achieve a given fixed $F_{eq}$ as a function of SV as CDFs. Uncertainties in the CDFs were obtained using bootstrap sampling~\cite{Efron1993, Kay2014} with replacement  using the 209-patient sample. The median $C_{tol}$ for each $F_{eq} \in (0.5, 0.8, 0.9)$ and its 68\% confidence range were computed.  Decreasing $C_{tol}$ with increasing SV indicates a decreased clinical effect, while increasing $C_{tol}$ suggests an enhanced clinical effect with alternative delineations. 

\begin{figure}[htbp]
\begin{center}
\includegraphics[width=\linewidth]{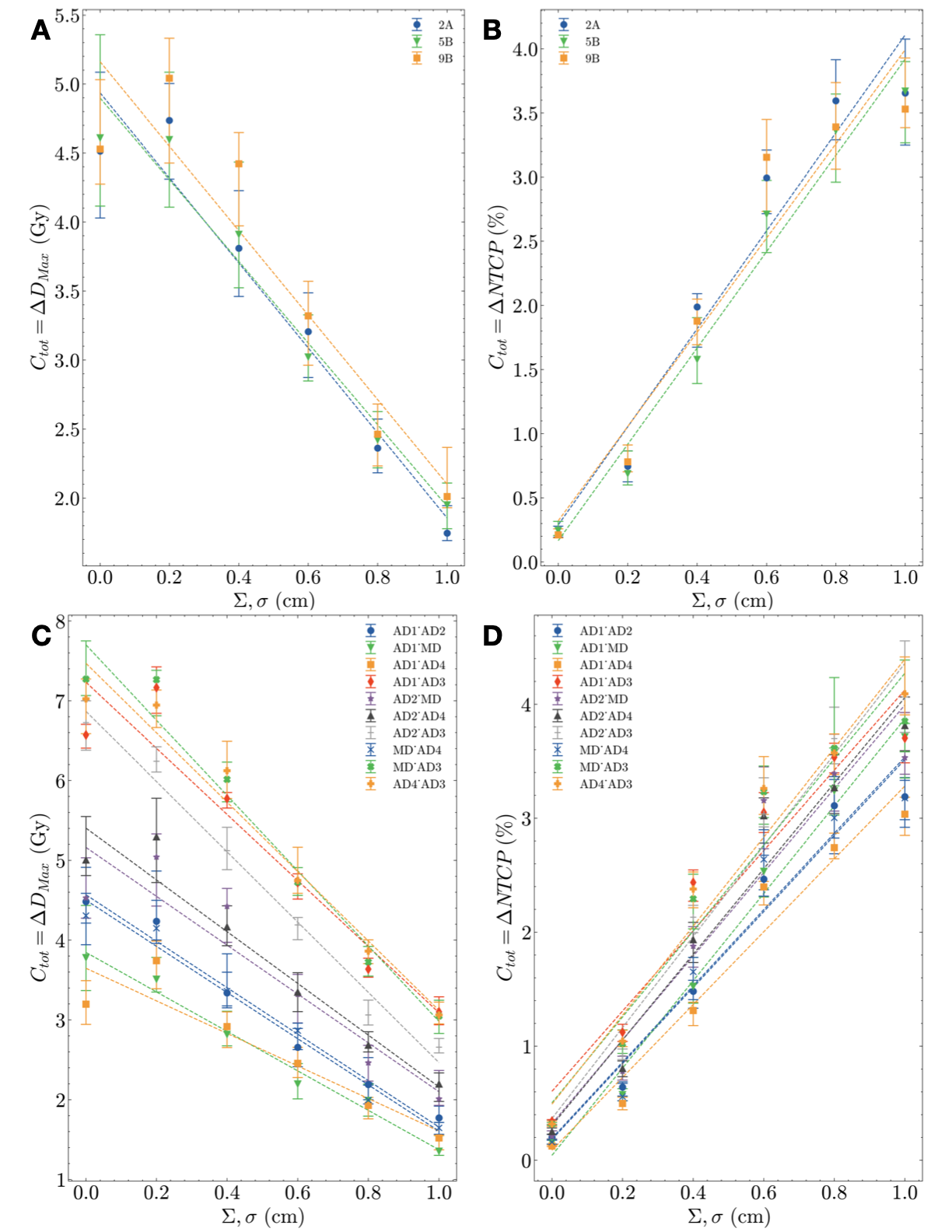}
\end{center}
\caption{$C_{tol}$ for $F_{eq} = 0.9$ for SpinalCord delineations with each simulated setup uncertainty level. Bootstrap sampling with 10000 iterations was used to account for the sampling uncertainty. The median $\Delta PQI$, and the 68\% confidence level are reported. Dotted lines represent the weighted linear fit. 
(A) and (B) compare different planning techniques with a fixed PD-TD combination (MD \& AD2) for $\Delta D_{\text{max}}$ and $\Delta NTCP$ respectively.
Different PD-TD combinations for a fixed planning technique (9B) are compared in (C) and (D).
}
\label{fig:MaxDelta90}
\end{figure}

Figure~\ref{fig:MaxDelta90} shows results for $F_{eq}=0.9$ for the 2-arc plans and 
SpinalCord $(PQI \in (D_{\text{max}}, NTCP)$. 
The negative slope in SV \emph{vs.} $C_{tol}$ for $D_{\text{max}}$ evaluations indicates a washout effect, consistent with Aliotta et al.~\cite{Aliotta2019}.
Conversely, the positive slope in the NTCP evaluations indicates increased clinical effect of DV with SV.

Figure~\ref{fig:MaxDelta90}(A) and (B) compare the 3 treatment planning beam arrangements for $F_{eq} = 0.9$ and SpinalCord. 
Increasing SV reduces the impact of DV for the DVH metric but it increases it for NTCP across all planning techniques. 
Panels (C) and (D) 
compare of the effect of SV 
for different PD-TD combinations for SpinalCord, showing similiar trends.
Differences measured using dose-based PQIs between PD-TD combinations are larger than those observed from NTCP for clinically relevant setup variabilities (0mm-4mm).

\section{Discussion}

While some previous studies found significant differences in dose-based $PQI$s evaluated on alternative delineations when evaluated on the static treatment plan~\cite{Fung2020, Guo2021},
when inherent SV is considered, Aliotta et al.~\cite{Aliotta2019} found that dose-based $PQI$ differences decreased as simulated SV increased. 
Our results align with these findings (except for non-target-overlapping parotids);  $\Delta D_{\text{mean}}$ and $\Delta D_{\text{max}}$ values from alternative delineations decreased with increasing  simulated SV. However, NTCP differences increases with SV (except for the Glnd\_Submands).

Previous studies reported weak correlation between GIs and dosimetric changes for alternative delineations~\cite{Lim2019, Fung2020, Guo2021}, indicating that GIs are inadequate for determining the clinical adequacy of delineations~\cite{Zhu2020}. Our findings support those conclusions.  
However, these conclusions are from prioritizing PTV coverage over OAR sparing. 
If OAR sparing was prioritized over PTV coverage (as in lung trials), then the sensitivity to geometric changes could be higher.\cite{Hoffmann2022}.

In preliminary testing, the auto-planning algorithm terminated with high (>55 Gy) SpinalCord  $D_{\text{max}}$ for some patients.  These plans, which would never be used clinically, had particularly high sensitivity of $\Delta NTCP$  to increasing SV even though $\Delta D_{\text{max}}$ reduced with increasing SV. This is due to the large slope of the sigmoidal NTCP curve at large $D_{\text{max}}$ values. Final auto-plans reduced the SpinalCord $D_{\text{max}}$ (to <55 Gy), and reduced, but did not eliminate this effect. 

To ensure the delineations used differed, extensive delineation review and adjustment by medical experts was not performed in this study, even though adjustment of AI contours improves their geometric conformance to manual delineations~\cite{Doolan2023}, the reported improvements are small (average vDSC improvement 0.02$\pm$0.02).  
Thus, our delineation variations may be greater than clinical practice, but this  is unlikely to affect our conclusions.

Delineations excluded for vDSC<0.5 had outlier $\Delta PQIs$ values and were often clearly erroneous (e.g. an AI SpinalCord miss-placed to the posterior skull).
Additionally, a few (MD SpinalCord) delineations were missing slices, which was corrected for by interpolation prior to planning.    

Visual inspection revealed systematic differences between some delineation sets.  For instance, one AD set SpinalCord encompassed the entire spinal canal, while other conformed to the SpinalCord.  Despite this, equivalence evaluations followed the same trends for this set.

Using common dose levels and auto-planning techniques may have added clinically conservative aspects to our study.  
While clinically, different dose levels are used based on the primary disease and nodal involvement,  we consistently used 70 Gy, yield higher doses to OARs.  Similarly, our lack of beam or collimator angle optimization for our 5-beam, 9-beam, and 2-arc VMAT plans also contributed to conservatively high OAR doses, as did the lack of post-auto--planning dosimetrist tuning to provide additional OAR protection.

Our dataset had a low fraction (8\%) of parotids not overlapping with target volumes. For these parotids, the $\Delta D_{\text{mean}, 95}$ increased as the simulated SV increased 
with SV, unlike overlapping parotids and other OARs.  
The non-overlap parotids SV dependence is due to the dose blurring-effect of random SV moving dose from the adjacent high-dose regions into the parotid.

Differences in SV effects between ($\Delta D_{\text{max}}$, $\Delta D_{\text{mean}}$) and $\Delta NTCP$ stresses the need to focus on clinical effects, rather than just dose-metrics.  Large dose-metric changes can be clinically inconsequential for NTCP, while small changes near dose-metric tolerance can significantly change NTCP. Cases where DV alone caused a dose metric violation occurred in less than 1\% of the cases studied.

\section{Conclusion}

Our study highlights the complex interplay between SV and DV in radiotherapy planning. We found that increasing SV generally reduces differences in dose-volume histogram (DVH) metrics between alternative delineations, suggesting a washout effect. Conversely, NTCP metrics show an increase in differences between delineations as SV rises. This pattern holds true across various treatment plans, including 5-Beam IMRT, 9-Beam IMRT, and 2-Arc VMAT, as well as across different delineator combinations.

The accuracy required for delineation is case-specific, influenced by factors beyond simple geometric variations. This underscores the need for personalized assessments in treatment planning to ensure optimal outcomes. Effective quality assurance (QA) programs must incorporate both geometrical variation analysis and dosimetric impact analysis to address the multifaceted challenges presented by delineation and setup variability in clinical practice.

Our findings also have significant implications for clinical workflows in radiotherapy. A nuanced understanding of how SV influences the clinical impact of DV necessitates careful evaluation and potential adjustment of treatment plans. Such adjustments are crucial to accommodate the varying degrees of SV and DV encountered in daily clinical practice.

In summary, this study provides pivotal insights into the complexities of the effects of delineations and delineation variation radiotherapy treatment planning. It emphasizes the importance of considering both setup and delineation variability in developing robust, effective, and personalized treatment strategies, to enhance the overall quality and efficacy of patient care in radiotherapy.

\section*{Acknowledgments}
This work was supported by National Cancer Institute (NCI) of the National Institute of Health under the award number R01CA222216. The content is solely the responsibility of the authors and does not necessarily represent the official views of the National Institutes of Health. We would like to thank Carina Medical, Radformation and Siemens Healthineers for providing us with AI delineations.

\bibliographystyle{plainurl}
\bibliographystyle{unsrt}
\end{document}